\documentclass[aip,rsi,superscriptaddress, preprint]{revtex4-1}
\usepackage{graphicx,CJK,times}

\begin{document}
\begin{CJK*}{UTF8}{}

\title{Robust and economical multi-sample, multi-wavelength UV/vis absorption and
fluorescence detector for biological and chemical contamination}

\author{Peter J.~Lu (\CJKfamily{bsmi}陸述義)}
\affiliation{Department of Physics and SEAS, Harvard University, Cambridge,
Massachusetts 02138, USA}

\author{Melanie M.~Hoehl}
\affiliation{Harvard-MIT Division of Health Sciences Technology, Cambridge MA 02139 USA}
\affiliation{Department of Mechanical Engineering, MIT, Cambridge MA 02139 USA}

\author{James B.~Macarthur}
\affiliation{Department of Physics and SEAS, Harvard University, Cambridge,
Massachusetts 02138, USA}

\author{Peter A.~Sims}
\affiliation{Department of Biochemistry and Molecular Biophysics, Columbia University
Medical Center, New York, New York, 10032, USA}
\affiliation{Columbia Initiative in Systems Biology, Columbia University Medical
Center, New York, New York, 10032, USA}

\author{Hongshen Ma}
\affiliation{Department of Mechanical Engineering, MIT, Cambridge MA 02139 USA}
\affiliation{Department of Mechanical Engineering, University of British
Columbia, Vancouver, British Columbia, Canada V6T 1Z4}

\author{Alexander H.~Slocum}
\affiliation{Department of Mechanical Engineering, MIT, Cambridge MA 02139 USA}

\begin{abstract}
We present a portable multi-channel, multi-sample UV/vis absorption and
fluorescence detection device, which has no moving parts, can operate wirelessly
and on batteries, interfaces with smart mobile phones or tablets, and has the
sensitivity of commercial instruments costing an order of magnitude more. We use
UV absorption to measure the concentration of ethylene glycol in water solutions
at all levels above those deemed unsafe by the United States Food and Drug
Administration; in addition we use fluorescence to measure the concentration of
\textit{d}-glucose. Both wavelengths can be used concurrently to increase
measurement robustness and increase detection sensitivity. Our small robust
economical device can be deployed in the absence of laboratory infrastructure,
and therefore may find applications immediately following natural disasters, and
in more general deployment for much broader-based testing of food, agricultural
and household products to prevent outbreaks of poisoning and disease.
\end{abstract}
\maketitle
\end{CJK*}

\section{Introduction}
Commercial laboratory spectrophotometers and fluorometers are particularly
useful for identifying the presence of unknown compounds, and for characterizing
the absorption, emission and/or fluorescence spectra over a broad range of
user-determined wavelengths, with great specificity and flexibility.
However, these instruments generally require a laboratory environment: the
mechanical components that enable the wide range of precisely-selectable
wavelengths, including movable diffraction gratings and lens assemblies, demand
mechanical and thermal stability, while the halogen lamps that generate UV
wavelengths used for absorption or fluorescence require significant electrical
power. As a result, these precision instruments are generally unsuitable for use
outside the laboratory. There are a number of handheld devices that are
starting to bridge the gap between laboratory and field analysis, but these are
still typically orders of magnitude too expensive for widespread field use where
they are needed to test, for example, food safety.

There are indeed many situations where the ability to measure absorption and
fluorescence outside of the laboratory can be critically important; for example,
detecting contaminants in food, medicine and agricultural products in the field,
or evaluating microbial contamination of drinking water in the immediate
aftermath of a natural disaster. In many countries, contamination testing is
simply not performed, on account of the cost of analytical laboratory equipment
and a scarcity of trained
users~\cite{Gomes_WHO_2002,Velusamy_BioTechAdv_2010,Yager_AnnRevBiomedEng_2008}.
In many of these cases, the specific wavelengths of interest to be probed are
\textit{a priori} known, and the user would simply like to measure the
absorption or emission, and / or the subsequent time evolution of these
quantities, at known wavelengths. For example, detecting contamination of known
compounds, or monitoring the progress of reactions involving known dyes or
fluorophores, are often-used applications where the spectra are known, but the
absolute measure of intensity and its time evolution are of interest. In many of
these cases, particularly those involving contamination, the ability to do
spectroscopy outside the laboratory environment and in the field can be of
tremendous value. By fixing wavelengths, the mechanical complexity of
instruments to detect absorption and fluorescence may be removed, opening up the
possibility of precise measurements outside the laboratory.

In this paper, we present a new device we created in response to the many cases
of DEG poisoning of children we read about because we felt as engineers we
should be able to create a robust affordable detection device whose very
deployment could potentially deter the substituion of DEG for gylcerine. Our
device combines multiple UV/vis absorption and fluorescence detectors, in a
geometry allowing simultaneous, multi-channel measurement of the time evolution
of multiple samples, with single-wavelength sensitivity comparable to laboratory
instruments that cost many orders of magnitude more. By combining the data from
two wavelengths, substances that may have a significant background measured at
one wavelength can benefit from data collected in the other, where the
background signal may be substantially less. This significantly increases the
signal-to-noise ratio in the measurement, and thus its effective sensitivity. To
test our hypothesis, we use UV absorption to monitor the progress of a
commercially-available enzyme-based kit to quantify the concentration of a major
chemical contaminant, ethylene glycol (EG), in water---which in the past has
episodically killed or sickened hundreds; we measure EG concentrations correctly
in water, at all levels above those deemed safe by the United States Food and
Drug Administration (FDA) and the European Community~\cite{FDA_2010,SCCP_2008}.
In addition, we use fluorescence to measure the concentration of glucose using a
simple dye-based enzyme kit, and find that the accuracy and sensitivity of our
device equals that of a commercial plate reader. Our device is completely
self-contained and has no moving parts, can run on batteries or solar panels,
can connect wirelessly, and can employ a mobile-phone / tablet platform for
computation, data analysis and network connectivity. Consequently, our device
can be deployed essentially independent of infrastructure---such as in remote or
disaster-stricken areas---where laboratory instruments simply cannot be used;
our technology might therefore provide key information to first-responders in
preventing further tragic episodes. Moreover, the rapid measuring time and very
affordable nature of our device and concomitant chemistry allow our system to be
widely deployed in a practical manner, potentially enabling substantially
broader systematic testing of foods, household products and medicines, the vast
majority of which are simply not tested at present~\cite{FDA_Rpt_2007}; thus,
our system may have the potential to decrease significantly the number of
poisoning incidents worldwide and save lives.

\section{Design and Fabrication}
Our overall design philosophy uses low-cost, commercially-available elements,
precisely positioned using molded-in-place reference features with respect to a
sample, to provide maximum mechanical and optical repeatability that then
simplifies data analysis, while minimizing complexity and therefore the cost of
manufacturing and deployment. By combining multiple UV and visible light sources
and detectors in close proximity to illuminate the sample, we achieve greater
detection robustness. Our design can be evolved easily to add an additional
light source such as an infrared source/detector, made possible by using a
cylindrical test tube, as opposed to a traditional rectangular cuvette, that is
then encircled by the sensors and detectors. Simple injection-molded parts allow
for the insertion of different thin plastic filters, LEDs and detectors in close
proximity to each other and the test tube. To cover a wide range of wavelengths,
traditional spectrophotometers and fluorometers use a broadband UV/vis light
source and movable diffraction gratings, costly precision components that are
mechanically fragile, and require constant maintenance and calibration. In our
device, because the spectra of the substances to be detected are \textit{a
priori} known, we can choose an illumination source that emits in a narrow range
of wavelengths. Lasers could be a convenient choice, but LEDs are generally more
rugged and compact, require less power, and are now available in a wide variety
of wavelengths from the deep UV to long-wavelength IR. Furthermore, they can be
extremely inexpensive, costing a few cents to dollars, orders of magnitude less
than the xenon lamp in a typical spectrophotometer. We choose a standard LED
size for illumination, which allows us to take advantage of the wide variety of
wavelengths available commercially---and any LED in the same form factor can be
substituted without modification to either the device electronics or mechanical
structure.

The spectra of some LEDs, though centered around a desired
wavelength, can have relatively broad tails. To improve the spectral specificity
of these LEDs for spectroscopy, we filter out undesired wavelengths with
body-colored polycarbonate films (Roscolux) designed for theatrical lighting,
which are very inexpensive and designed for durability in high-heat and light
environments; the small pieces we use in our device cost fractions of a cent.

\begin{figure}
\includegraphics[width=5in]{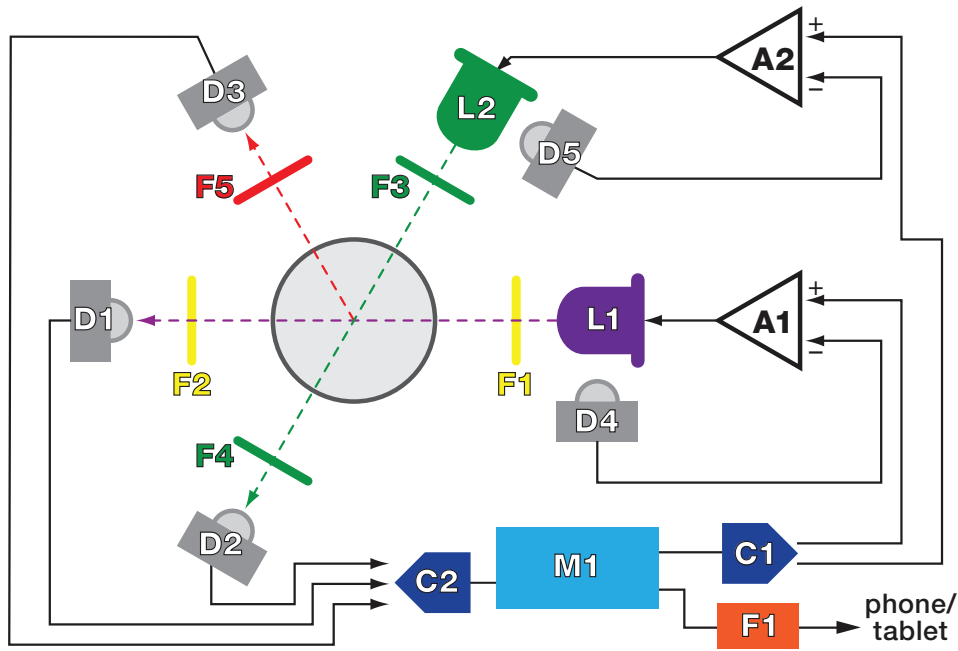}
\caption{
Device schematic. UV light emitted by an LED (L1) passes through an excitation
filter (F1), the sample, and another filter (F2) before its absorption is
detected (D1). Optical feedback using an additional sensor (D4) and op-amp (A1)
maintains a constant light output from L1, whose level is set by the
microcontroller (M1) via a voltage generated by a D/A converter (C1). Light from
a similarly stabilized green LED (L2, D5, A2) is filtered (F3) before passing
through the sample; green light is filtered and detected for green absorption
(F4, D2) and red fluorescence (F5, D3). Voltage outputs from the detectors (D1,
D2, D3) are digitized by an A/D converter (C2) and sent to the microcontroller
(M1), which formats and transmits the data via USB (F1) to a computer, smart
mobile-phone or tablet.}
\label{fig_device_schematic}
\end{figure}

\begin{figure}
\includegraphics[width=4in]{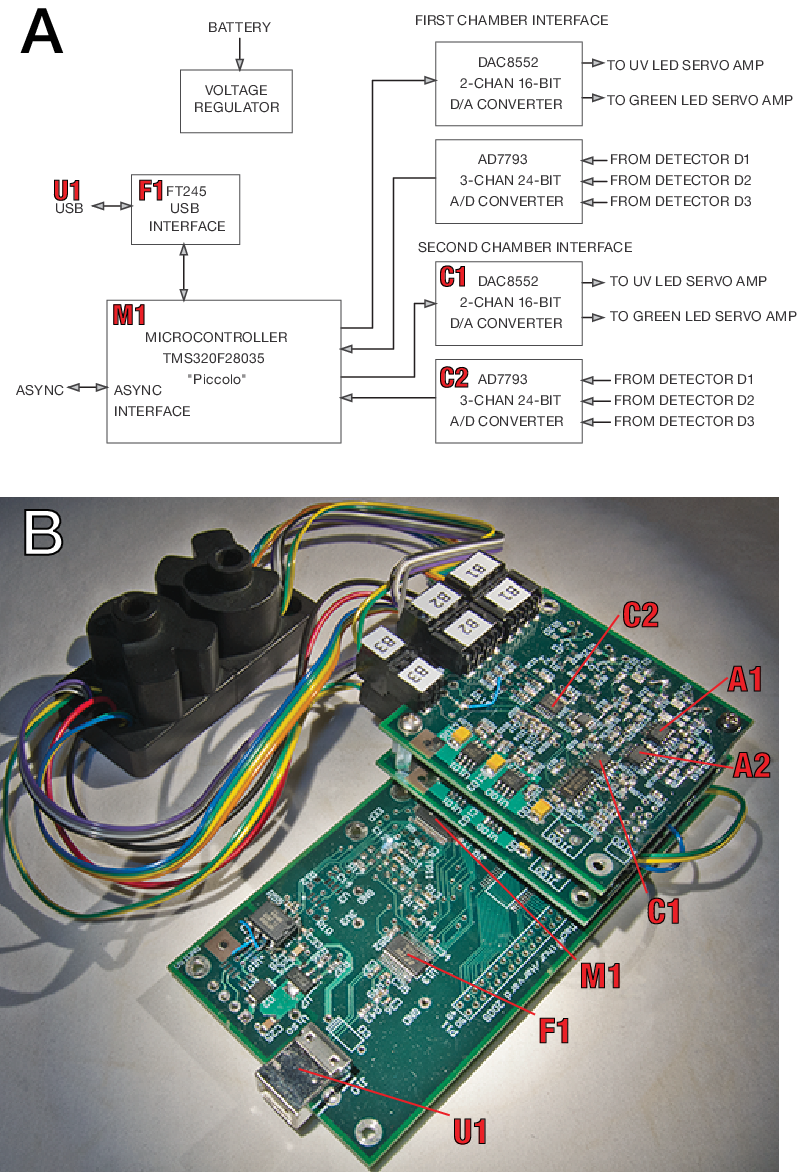}
\caption{
(A) Circuit diagram showing specific electrical components of our detector, with
microcontroller (M1), D/A converter (C1), A/D converter (C2) and
USB Interface (F1) components labeled as in Fig.~\ref{fig_device_schematic}. (B)
Photograph of the detector (black, upper left) and circuit boards, with major
components labeled with red letter corresponding to labels in (A) and in
Fig.~\ref{fig_device_schematic}.}
\label{fig_circuit_diagram}
\end{figure}

The general geometry for absorption detection is shown in the device schematic
in Fig.~\ref{fig_device_schematic}. To measure absorption, we illuminate the
liquid sample with an LED (L1) and measure the
intensity change after the light has passed through the sample, using a photodiode. For simplicity and
robustness, we choose a form of photodiode integrated with an op-amp in a single
package (D1), which outputs a voltage proportional to the incident light
striking the photodiode (Texas Advanced Optoelectronic Solutions TSL257); this
semiconductor light-to-voltage detector costs around a dollar. The low cost of
this detector component allows us not only to use it for the primary absorption
measurement, but also to include a second detector (D4) next to the LED; we
couple it to an active feedback loop with an op-amp (A1) to stabilize the LED's
intensity at a constant level even as temperature changes, due to the external
environment or as the LED is powered on. The circuit is a single-amplifier
proportional servo with a bandwidth of 1 KHz, which effectively removes thermal
drift and reduces errors from mechanical vibration. The high-level voltage
output of the sensor makes for a simplified servo circuit---each LED's driver
adjusts the LED current until the sensor voltage equals the voltage set by the
microcontroller (M1) via a D/A converter (C1). This active illumination
stabilization, a feature found in advanced laser systems, is crucial to
reproducible light intensity measurements. The analog voltage from the detector
is converted via an A/D converter (C2) to a digital value that is sent to a
microcontroller (M1), then as ASCII over wired or wireless USB (U1) to a
Linux-based (Ubuntu) host, where the data processing and analysis (C++), storage
and communication are performed. The output data is a simple list of 16-bit
integers, corresponding to the digitized voltage levels from each
light-to-voltage detector; 16 bits are sufficient to capture all of the signal
above the noise from the detectors, but we use 24-bit detectors because present
discrete inexpensive A/D converters all use this higher resolution. We use a
variety of host computing platforms, including an ordinary Intel-based laptop,
and an ARM-based mobile phone / tablet platform (NVIDIA Tegra family), with
equivalent functionality. A circuit diagram with specific components is shown in
Fig.~\ref{fig_circuit_diagram}A; a photograph of the circuit boards and the
device itself is shown in Fig.~\ref{fig_circuit_diagram}B.

\begin{figure}
\includegraphics[width=4in]{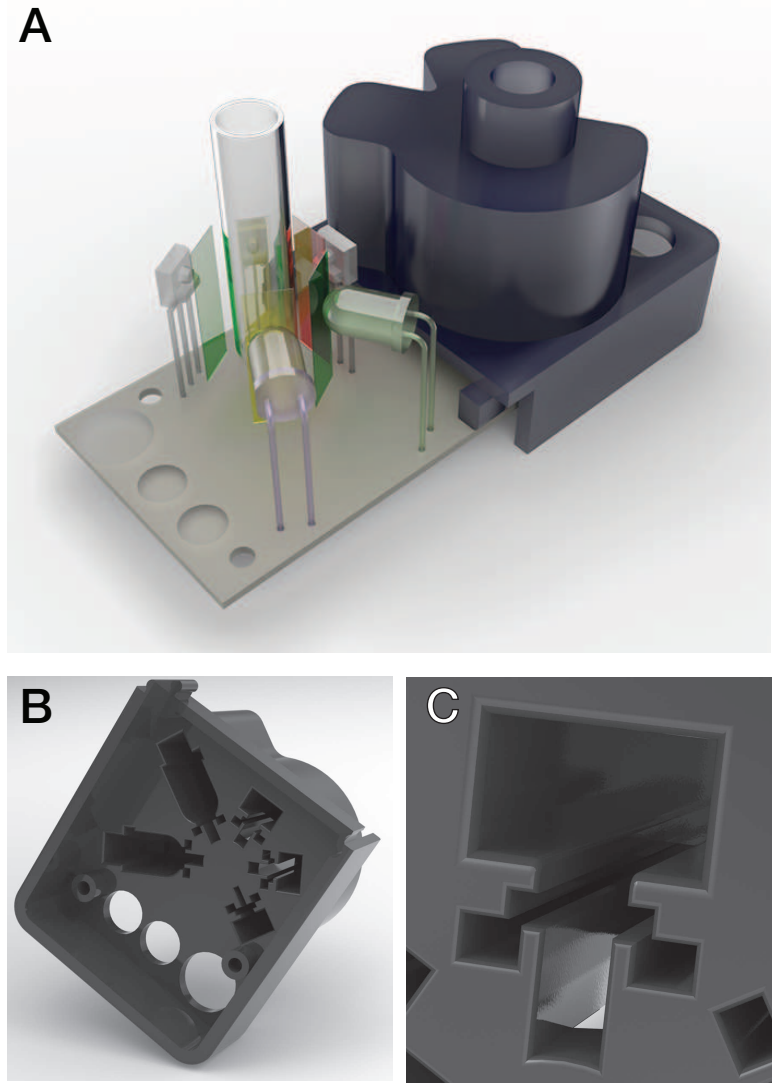}
\caption{
(A) 3D model of our as-built device. One chamber has the black plastic enclosure
removed, to illustrate the positions of LEDs, filters and sample-holding test
tube. (B) View of the bottom of the plastic enclosure, showing openings for two
LEDs (L1 and L2 in Fig.~\ref{fig_device_schematic}) and 3 light-to-voltage
detectors (D1, D2 and D3 on Fig.~\ref{fig_device_schematic}). (C) Close-up of
the opening for one light-to-voltage detector, showing slit for thin filter
plastic.}
\label{fig_device_rendering}
\end{figure}

The LED, sample, detector and filters are held in place deterministically by an
opaque plastic enclosure that is extremely simple to manufacture and robust in
use. The actual device appears in the photograph in
Fig.~\ref{fig_circuit_diagram}B. A 3D model of the device, raytraced on an
NVIDIA Tesla C2050 GPU using Bunkspeed Pro, with one of the two enclosures
removed to show the position of the LEDs and detectors, is shown in
Fig.~\ref{fig_device_rendering}A. This monolithic structure is designed for
two-part mold injection molding with no side pulls; our experimental unit is
made by stereolithography (SLA), and in production would cost less than a
dollar. We achieve high-precision location of the LED, sample, and detectors by
using the components' leads themselves as flexure spring features, which force
the components against reference features, all within the same monolithic part.
The bottom of the enclosure is shown in Fig.~\ref{fig_device_rendering}B,
showing a clear view of the cavities to hold the components; a close-up of one
of the light-to-voltage detector cavities in shown in
Fig.~\ref{fig_device_rendering}C. A key to successful optical detection is to
create a very thin slit to receive the plastic filter without allowing any light
leakage. To mold such a slit using a thin core would not be practical, so
instead larger near-overlapping mold cores are used, such that when the mold is
opened, an effective thin gap is created, as shown in
Fig.~\ref{fig_device_rendering}C.

To enable the sample to be illuminated from multiple sources in a small space,
we use a round glass test tube (Durham 6$\times$50 culture tube) as our chemical
reaction chamber, whose hemispherical bottom rests in a hemispherical seat to
establish axial and radial position; clearance around the test tube walls
accommodates tolerances, while use of a test tube with length-to-diameter ratio
greater than ten keeps the tilt of the test tube to a level that does not affect
the readings. Our mechanically-robust design has no moving parts and ensures
positional repeatability; samples removed, reinserted, or measured in a
different test tubes, all yield the same results.

\section{Preliminary Experiments and Results}
We conduct an initial series of tests to validate the design and enable us to
plan for further measurements. We first test the UV absorption capabilities of
our device with an enzyme-based assay that detects the presence of ethylene
glycol (EG), an important component of automotive antifreeze that is sometimes
improperly substituted for propylene glycol and polyethylene glycol in
medicines, household products and foods, with lethal consequences\footnote{EG
poisonings episodically kill hundreds in third-world countries, even today. The
realization that there is no existing low-cost method for detecting such poisons
motivated the research that led to the device and chemistry methods presented
here.}. Within the human body, EG is first digested by alcohol dehydrogenase
(ADH) into an aldehyde, then into various acids, which are the direct cause of
physiological damage~\cite{Gomes_WHO_2002}. To detect EG, we choose an ADH-based
enzyme assay, which converts a hydroxyl group to an aldehyde, and at the same
time converts the coenzyme NAD$^+$ into NADH~\cite{Eckfeldt_ClinChem_1980}. NADH
has a broad absorption peak from 350-370 nm, so that the measured absorption of
UV light in that wavelength range should reflect the amount of EG; we therefore
select a UV LED (ledsupply.com L5-0-U5TH15-1) with a peak emission of
$\lambda=365$ nm, to illuminate the sample.

\begin{figure}
\includegraphics[width=4in]{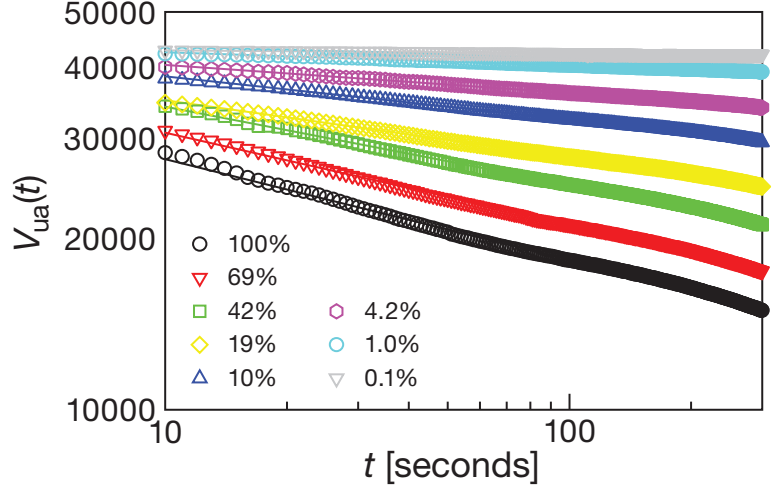}
\caption{
Time evolution of output voltage $V_\mathrm{ua}(t)$ from the UV detectors,
digitized as 16-bit integer, shown on a log-log plot with symbols for different
EG concentrations $c_\epsilon$ in water. The data fall onto a straight line for
each sample, demonstrating power-law scaling. The magnitude of the slope of each
line $\gamma(c_\epsilon)$ varies monotonically with $c_\epsilon$.}
\label{fig_UV_voltage_time}
\end{figure}

To measure $c_\epsilon$, the mass fraction (concentration) of EG, we add a
solution of ADH to the sample, immediately insert into the sample chamber, and
record the voltage $V_\mathrm{ua}(t, c_\epsilon)$ measured by the UV absorption
detector once per second for five minutes. For pure EG ($c_\epsilon=1$), the
$V_\mathrm{ua}(t, c_\epsilon)$ data fall on a straight line when plotted on a
log-log plot, demonstrating a power-law behavior, as shown by the black circles
in Fig.~\ref{fig_UV_voltage_time}. Because our test tube has a circular cross
section and the LED has a distribution of illumination angles, a single
path-length is not well-defined; therefore, we cannot rely on a simple Beer's
Law calculation for the absolute absorbance, particularly as there may be slight
lensing effects. Instead, we measure samples of known $c_\epsilon$ in water,
from the FDA safety limit of $c_\epsilon=10^{-3}$ to
$c_\epsilon=1$~\cite{FDA_2010}. This allows us to establish a calibration curve,
accounting for the slight variations in position and brightness / sensitivity of
the electronic components that will naturally vary from detector to detector. In
all cases we observe lines on the log-log plot; $V_\mathrm{ua}(t,c_\epsilon)
\propto t^{-\gamma(c_\epsilon)}$, as shown with colored symbols in
Fig.~\ref{fig_UV_voltage_time}.

\begin{figure}
\includegraphics[width=4in]{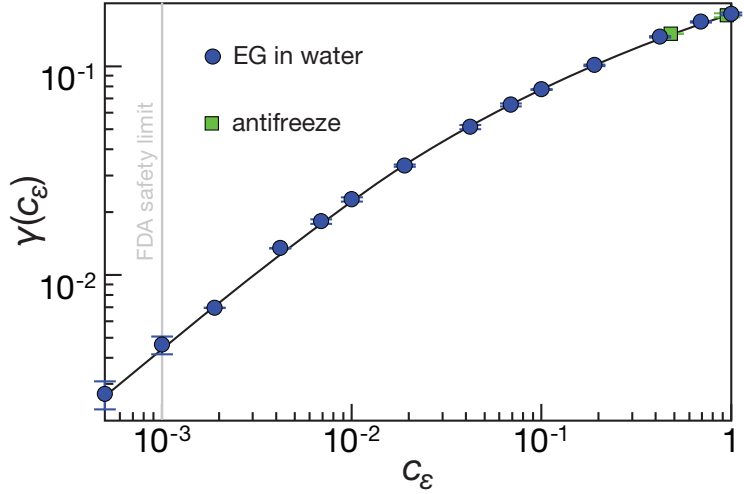}
\caption{
$\gamma(c_\epsilon)$ varies monotonically with $c_\epsilon$, shown with blue
circles for pure EG in water, and green squares for antifreeze, which follow the
same trend, shown in black as a guide to the eye. Data point represent averages
over several runs, with error bars corresponding to the standard deviation of
the measurements. FDA safety limit $c_\epsilon=10^{-3}$ is indicated with a grey
vertical line.}
\label{fig_UV_EG_conc}
\end{figure}

The power-law exponent magnitudes $\gamma(c_\epsilon)$ monotonically increase
with $c_\epsilon$, as shown with the blue circles in Fig.~\ref{fig_UV_EG_conc}.
The optical feedback loop stabilization (L1, D4 and A1 in
Fig.~\ref{fig_device_schematic}) ensures that LED intensity remains constant
irrespective of environmental changes; thus, there are no adjustable parameters
in our determination of $\gamma(c_\epsilon)$. These data demonstrate our ability
to measure $c_\epsilon$ in drinking water, which has caused sickness and death
even in the United States~\cite{Schultz_MMWR_1987}, at all concentrations deemed
unsafe by the FDA. To confirm the absolute, not just relative, accuracy of our
data, we also measure antifreeze samples with known EG concentrations, following
the same chemical protocol. Strikingly, we find that these samples fall on the
same curve as the EG in water, shown by the green squares in
Fig.~\ref{fig_UV_EG_conc}. These data demonstrate that our measurements may
provide a quantitatively accurate way to measure absolute EG concentration.

In addition to absorption, fluorescence is another common spectroscopic
technique. The use of different filters allows us to excite at one wavelength,
and detect emission at a longer one. The circular geometry of the system, with
the light-tight filter-holding features, allows us to integrate a second,
separate fluorescence channel for the same sample as the absorption detection
channel without any interference. Hence we add another LED spaced 60$^\circ$
from the UV LED for excitation, and two additional light detectors, using
differently-colored theater gel plastic to filter the green absorption and red
fluorescence, placed at 180$^\circ$ and 60$^\circ$, respectively, relative to
the green LED. We choose these angles to accomodate all
of the LEDs and detectors, as using 90$^\circ$ square geometry does not allow
enough detectors to capture both UV absorption and visible fluorescence
simultaneously. The complete illumination and detection scheme is shown in
Fig.~\ref{fig_device_schematic}. We choose a green LED (Cree LC503FPG1-15P-A3)
with a peak emission of $\lambda=527$ nm for illumination in this
fluorescence channel, as there are a number of common red dyes that are excited
by light in this range; however, as in the case of the UV LED, any color can be
substituted within the same housing with the same electronics.

By offsetting the activation of the UV and green LEDs by 0.5 seconds, we collect
data from both UV-illuminated and green-illuminated channels once each second,
with no possible crosstalk. Thus, our device simultaneously measures absorption
and fluorescence with two excitation wavelengths and has no moving parts---which
is not possible with the square cuvette geometry traditionally found in
laboratory fluorometers and spectrophotometers. Moreover, the low cost of our
detector and electronics makes it economical for us to construct a second,
identical, multi-channel detector. This provides the capability to run control
reactions in parallel, so that the absolute activity of particular enzyme
samples can be divided out; the two-chamber final device is shown in
Figs.~\ref{fig_circuit_diagram}B and \ref{fig_device_rendering}A.

\begin{figure}
\includegraphics[width=4in]{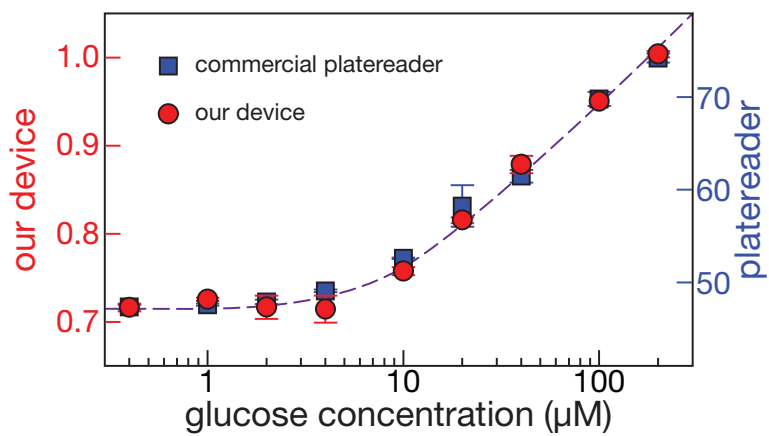}
\caption{Comparison of the detection of the same glucose oxidase-based reaction,
which generates a fluorescent product, in our detector and in a commercial
fluorscence plate reader. Data point represent averages over several runs,
with error bars corresponding to the standard deviation of the measurements.}
\label{fig_fluor_glucose}
\end{figure} 

To test the capability of our fluorescence detector, we measure the
concentration of glucose with a commercial kit based on the activity of glucose
oxidase (Invitrogen Amplex Red Glucose / Glucose Oxidase Assay Kit A22189). In
this assay, glucose oxidase reacts with $d$-glucose to form $d$-gluconolactone
and H$_2$O$_2$. In the presence of horseradish peroxidase (HRP), the H$_2$O$_2$
reacts with the Amplex Red reagent in a 1:1 stoichiometry to generate a red
fluorescent oxidation product, related to resorufin, which emits a bright red
color when excited with green light. We prepare the various enzyme and dye
solutions following the manufacturer's protocol, and run the reaction on samples
of various concentrations of \textit{d}-glucose, ranging from 0.4 $\mu$M to 400
$\mu$M. The UV absorption and green fluorescence values are monitored in the
detector for 5 minutes and again measured after 30 minutes to provide a
stationary fluorescence value after the reaction is complete. At the same time
as each sample is run, we also run a second control sample with a
\textit{d}-glucose concentration of 200 $\mu$M in the second chamber, whose
value we divide out in order to cancel the effects of variations with enzyme
activity. This normalization, made possible by the presence of the second
detector, allows us to quantify the amount of $d$-glucose at all concentrations
greater than few $\mu$M, as shown in Fig.~\ref{fig_fluor_glucose}.

We compare the sensitivity of our device to that available in commercial
instruments, running the same set of samples with the same enzyme kit in a
fluorescence plate reader (Molecular Devices, SPECTRAmaxTM GEMINI XS), exciting
at 530 nm, and detecting at 590 nm, again after 30 minutes. Our device has the
same lower-limit of detectable $d$-glucose concentrations as the commercial
plate reader, which costs several orders of magnitude more, as shown in
Fig.~\ref{fig_fluor_glucose}.

\section{Conclusions and Further Work}
The capabilities of our device make it suitable for a broad range of
applications where spectroscopic wavelengths are \textit{a priori} known. The
individual channels within our detector have sensitivity comparable to
commercial laboratory instruments costing orders of magnitude more; moreover,
the low cost of our device makes practical the multiple channels and samples
that allow us to normalize by references and combine data from different
simultaneous measurements to remove background and substantially increase the
signal-to-noise ratio. Our device is self-contained, in contrast to the array of
pumps and additional off-chip infrastructure needed to support a microfluidic
device. It consumes sufficiently low power to run on batteries or a solar panel,
is mechanically rugged and functions in the absence of external infrastructure;
hence our device has potential application in remote areas and disaster
situations where even the most basic measurements have the potential to save
many lives.

Widespread deployment of the device we have presented, or variants thereof,
coupled with the network connectivity and geo-location technology present in the
smart mobile-phone / tablet platform we use for analysis and computation, can
facilitate automated aggregation of data remotely; we can imagine that
geographic and temporal patterns, for example, in detected contamination may
allow earlier identification of outbreaks than is now possible, adding a new
tool to control and lessen the effects of epidemics.
Finally, the fast detection of our assays---only a few minutes---allows rapid
testing of perishable foods and ingestible products which are not tested because
current culturing-based methods require days; this new capability may have
potential applications in much broader sampling of both domestic and imported
foods and agricultural products, enabling for the first time end-to-end
characterization within a food or medicine supply chain.

\begin{acknowledgements} 
We thank J.~Sims, J.~Helferich, I.~Wong, S.~Finch, J.~Voldmann, Y.~Wang,
D.~Chait and B.~Stupak. This work supported by CIMIT, the Legatum Center at
MIT, the NVIDIA Professor Partnership Program, and personal funds of the
authors.
\end{acknowledgements}

\begin{thebibliography}{9}%
\makeatletter
\providecommand \@ifxundefined [1]{%
 \@ifx{#1\undefined}
}%
\providecommand \@ifnum [1]{%
 \ifnum #1\expandafter \@firstoftwo
 \else \expandafter \@secondoftwo
 \fi
}%
\providecommand \@ifx [1]{%
 \ifx #1\expandafter \@firstoftwo
 \else \expandafter \@secondoftwo
 \fi
}%
\providecommand \natexlab [1]{#1}%
\providecommand \enquote [1]{``#1''}%
\providecommand \bibnamefont [1]{#1}%
\providecommand \bibfnamefont [1]{#1}%
\providecommand \citenamefont [1]{#1}%
\providecommand \href@noop [0]{\@secondoftwo}%
\providecommand \href [0]{\begingroup \@sanitize@url \@href}%
\providecommand \@href[1]{\@@startlink{#1}\@@href}%
\providecommand \@@href[1]{\endgroup#1\@@endlink}%
\providecommand \@sanitize@url [0]{\catcode `\\12\catcode `\$12\catcode
 `\&12\catcode `\#12\catcode `\^12\catcode `\_12\catcode `\%12\relax}%
\providecommand \@@startlink[1]{}%
\providecommand \@@endlink[0]{}%
\providecommand \url [0]{\begingroup\@sanitize@url \@url }%
\providecommand \@url [1]{\endgroup\@href {#1}{\urlprefix }}%
\providecommand \urlprefix [0]{URL }%
\providecommand \Eprint [0]{\href }%
\providecommand \doibase [0]{http://dx.doi.org/}%
\providecommand \selectlanguage [0]{\@gobble}%
\providecommand \bibinfo [0]{\@secondoftwo}%
\providecommand \bibfield [0]{\@secondoftwo}%
\providecommand \translation [1]{[#1]}%
\providecommand \BibitemOpen [0]{}%
\providecommand \bibitemStop [0]{}%
\providecommand \bibitemNoStop [0]{.\EOS\space}%
\providecommand \EOS [0]{\spacefactor3000\relax}%
\providecommand \BibitemShut [1]{\csname bibitem#1\endcsname}%
\let\auto@bib@innerbib\@empty
\bibitem [{\citenamefont {Gomes}, \citenamefont {Liteplo},\ and\ \citenamefont
 {Meek}(2002)}]{Gomes_WHO_2002}%
 \BibitemOpen
 \bibfield {author} {\bibinfo {author} {\bibfnamefont {R.}~\bibnamefont
 {Gomes}}, \bibinfo {author} {\bibfnamefont {R.}~\bibnamefont {Liteplo}}, \
 and\ \bibinfo {author} {\bibfnamefont {M.~E.}\ \bibnamefont {Meek}},\
 }\href@noop {} {\emph {\bibinfo {title} {Ethylene glycol: human health
 aspects}}}\ (\bibinfo {publisher} {World Health Organization},\ \bibinfo
 {year} {2002})\BibitemShut {NoStop}%
\bibitem [{\citenamefont {Velusamy}\ \emph {et~al.}(2010)\citenamefont
 {Velusamy}, \citenamefont {Arshak}, \citenamefont {Korostynska},
 \citenamefont {Oliwa},\ and\ \citenamefont
 {Adley}}]{Velusamy_BioTechAdv_2010}%
 \BibitemOpen
 \bibfield {author} {\bibinfo {author} {\bibfnamefont {V.}~\bibnamefont
 {Velusamy}}, \bibinfo {author} {\bibfnamefont {K.}~\bibnamefont {Arshak}},
 \bibinfo {author} {\bibfnamefont {O.}~\bibnamefont {Korostynska}}, \bibinfo
 {author} {\bibfnamefont {K.}~\bibnamefont {Oliwa}}, \ and\ \bibinfo {author}
 {\bibfnamefont {C.}~\bibnamefont {Adley}},\ }\href@noop {} {\bibfield
 {journal} {\bibinfo {journal} {Biotech.~Adv.}\ }\textbf {\bibinfo {volume}
 {28}},\ \bibinfo {pages} {232} (\bibinfo {year} {2010})}\BibitemShut
 {NoStop}%
\bibitem [{\citenamefont {Yager}, \citenamefont {Domingo},\ and\ \citenamefont
 {Gerdes}(2008)}]{Yager_AnnRevBiomedEng_2008}%
 \BibitemOpen
 \bibfield {author} {\bibinfo {author} {\bibfnamefont {P.}~\bibnamefont
 {Yager}}, \bibinfo {author} {\bibfnamefont {G.~J.}\ \bibnamefont {Domingo}},
 \ and\ \bibinfo {author} {\bibfnamefont {J.}~\bibnamefont {Gerdes}},\
 }\href@noop {} {\bibfield {journal} {\bibinfo {journal}
 {Annu.~Rev.~Biomed.~Eng.}\ }\textbf {\bibinfo {volume} {10}},\ \bibinfo
 {pages} {107} (\bibinfo {year} {2008})}\BibitemShut {NoStop}%
\bibitem [{\citenamefont {{United States Food and Drug
 Administration}}(2010)}]{FDA_2010}%
 \BibitemOpen
 \bibfield {author} {\bibinfo {author} {\bibnamefont {{United States Food and
 Drug Administration}}},\ }\href@noop {} {\emph {\bibinfo {title} {Guidance
 for Industry: Testing of Glycerin for Diethylene Glycol}}} (\bibinfo {year}
 {2010})\BibitemShut {NoStop}%
\bibitem [{\citenamefont {{European Commission Scientific Committee on Consumer
 Products}}(2008)}]{SCCP_2008}%
 \BibitemOpen
 \bibfield {author} {\bibinfo {author} {\bibnamefont {{European Commission
 Scientific Committee on Consumer Products}}},\ }\href@noop {} {\emph
 {\bibinfo {title} {Opinion on Diethylene Glycol}}} (\bibinfo {year}
 {2008})\BibitemShut {NoStop}%
\bibitem [{\citenamefont {{United States House or Representatives Subcommittee
 on Science and Technology}}(2007)}]{FDA_Rpt_2007}%
 \BibitemOpen
 \bibfield {author} {\bibinfo {author} {\bibnamefont {{United States House or
 Representatives Subcommittee on Science and Technology}}},\ }\href@noop {}
 {\emph {\bibinfo {title} {{FDA}: Science and Mission at Risk}}} (\bibinfo
 {year} {2007})\BibitemShut {NoStop}%
\bibitem [{Note1()}]{Note1}%
 \BibitemOpen
 \bibinfo {note} {EG poisonings episodically kill hundreds in third-world
 countries, even today. The realization that there is no existing low-cost
 method for detecting such poisons motivated the research that led to the
 device and chemistry methods presented here.}\BibitemShut {Stop}%
\bibitem [{\citenamefont {Eckfeldt}\ and\ \citenamefont
 {Light}(1980)}]{Eckfeldt_ClinChem_1980}%
 \BibitemOpen
 \bibfield {author} {\bibinfo {author} {\bibfnamefont {J.~H.}\ \bibnamefont
 {Eckfeldt}}\ and\ \bibinfo {author} {\bibfnamefont {R.~T.}\ \bibnamefont
 {Light}},\ }\href@noop {} {\bibfield {journal} {\bibinfo {journal}
 {Clin.~Chem.}\ }\textbf {\bibinfo {volume} {26}},\ \bibinfo {pages} {1278}
 (\bibinfo {year} {1980})}\BibitemShut {NoStop}%
\bibitem [{\citenamefont {Schultz}\ \emph {et~al.}(1987)\citenamefont {Schultz}
 \emph {et~al.}}]{Schultz_MMWR_1987}%
 \BibitemOpen
 \bibfield {author} {\bibinfo {author} {\bibfnamefont {S.}~\bibnamefont
 {Schultz}} \emph {et~al.},\ }\href@noop {} {\bibfield {journal} {\bibinfo
 {journal} {Morb.~Mortal.~Wkly.~Rep.}\ }\textbf {\bibinfo {volume} {36}},\
 \bibinfo {pages} {611} (\bibinfo {year} {1987})}\BibitemShut {NoStop}%
\end{thebibliography}
%

\end{document}